\documentclass[12pt]{iopart}
\usepackage[pdftex]{graphicx} 
\usepackage{xcolor}
\begin{document}

\title[Coherent coupling of single molecules to on-chip microresonators]{Coherent coupling of single molecules to on-chip ring resonators}

\author{Dominik Rattenbacher*$^1$, Alexey Shkarin*$^1$, Jan Renger$^1$, Tobias Utikal$^1$, Stephan G\"otzinger$^{2,3,1}$ Vahid Sandoghdar$^{1,2}$ }

\address{$^1$ Max Planck Institute for the Science of Light (MPL), Staudtstr. 2, 91058 Erlangen, Germany \\
$^2$ Department of Physics, Friedrich Alexander University Erlangen-N\"urnberg (FAU),
91058 Erlangen, Germany \\
$^3$ Graduate School in Advanced Optical Technologies (SAOT), FAU, 91052 Erlangen, Germany}
\ead{vahid.sandoghdar@mpl.mpg.de}
\vspace{10pt}

\begin{abstract}
We report on cryogenic coupling of organic molecules to ring microresonators obtained by looping sub-wavelength waveguides (nanoguides). We discuss fabrication and characterization of the chip-based nanophotonic elements which yield resonator finesse in the order of 20 when covered by molecular crystals. Our observed extinction dips from single molecules reach 22\%, consistent with the expected Purcell enhancements up to 11 folds. Future efforts will aim at efficient coupling of a handful of molecules via their interaction with a ring microresonator mode, setting the ground for the realization of quantum optical cooperative effects.

\end{abstract}

\section{Introduction}

Controlled coupling of photons to a large network of optical emitters is not only a crucial step for quantum engineering platforms such as quantum simulation \cite{Noh2017}, but it also provides an avenue for studying fundamental collective phenomena in quantum mechanics \cite{Chang2018}. A key requirement for achieving these goals is an efficient coupling of individual quantum emitters to a single photonic mode, which can be used as a ``wire" for interconnecting them \cite{Haakh2016}. 

Theoretically, a photon can be reflected with unity probability from a dipolar two-level emitter if its spatial mode is matched to that of the emitter's radiation pattern \cite{Zumhofen2008}. Indeed, substantial coupling efficiencies have been reported using various configurations such as near-field coupling \cite{Gerhardt2007}, free-space focusing \cite{Wrigge2008,Maser2016} and coupling to subwavelength waveguides (nanoguides) \cite{Faez2014,Türschmann2017,Javadi2018,Lombardi2018}. However, solid-state emitters confront a fundamental limitation in these endeavors because of their complex transition pathways and the influence of phonons. As a result, coherent emitter-photon interactions are hampered by a finite branching ratio defined as $\alpha=\Gamma_{\rm ZPL}/\Gamma_{\rm tot}$, where $\Gamma_{\rm ZPL}$ denotes the width of the zero-phonon-line (ZPL), and $\Gamma_{\rm tot}$ is the total linewidth. To remedy this issue, one can increase $\alpha$ by coupling the ZPL to a single mode of an optical microcavity, thus, rendering the effect of unwanted transitions insignificant \cite{Arcari2014,Sipahigil2016,Wang2019}. 

Very recently, we demonstrated near-unity coupling of single photons to the ZPL of an organic dye molecule placed in a high-finesse scannable Fabry-Perot microcavity \cite{Wang2019}. Although the open structure of such a cavity ensures a high degree of control on the cavity resonance frequency and facilitates optimization of the molecule position, its design is not easily scalable to multiple emitters. An elegant alternative would be to employ a chip-based architecture consisting of a network of nanoguides and  microresonators. This paper presents a realization of this goal by coupling single dibenzoterrylene (DBT) molecules embedded in a para-dichlorobenzene (pDCB) crystal to chip-based ring mircoresonators.

\section{Coupling to nanoguide ring-resonators}

It is known that the density of modes for the electromagnetic field is modified in the presence of boundaries, e.g., the mirrors of a Fabry-Perot cavity. This modification in turn affects the radiation properties of an emitter placed in it as compared to free space. The rate of emission into a single cavity mode ($\Gamma_\mathrm{cav}$) relative to that in free space ($\Gamma_0$) is given by the Purcell factor according to
\begin{equation}
F_P=\frac{\Gamma_\mathrm{cav}}{\Gamma_0}=\frac{3 \lambda^3}{4\pi^2\epsilon^{3/2}} \frac{Q}{V}\,,
\label{Purcell}
\end{equation}
where $\lambda$ is the transition wavelength in vacuum, $\epsilon$ denotes the permittivity, $Q$ is the quality factor, and $V$ signifies the cavity mode volume \cite{Walther2006}. Thus, large enhancements can arise from high $Q$s and low $V$s, i.e. if the mode is strongly confined in space and in frequency. 

In our current work, we use a microscopic ring resonator composed of a looped nanoguide evanescently coupled to two other nanoguides, as illustrated in figure\,\ref{schematics}(a). In order to describe the coupling of a dipolar emitter to this structure, it is instructive to consider the emission into a linear nanoguide first. This can be calculated from the knowledge of the classical Green's tensor, which accounts for the geometric and material properties of the nanoguide and its surrounding \cite{Haakh2016,Welsch1996,Garcia2017b}. The Green's tensor for propagation in a linear nanoguide can be decomposed into a transverse component ($G_\mathrm{\perp}$) and a part along the nanoguide ($G_\mathrm{1D}$), yielding $G=G_\mathrm{\perp}G_\mathrm{1D}$, where $G_{\mathrm{1D}}(k(\omega),z,z^\prime)=\frac{i}{2k}e^{ik \vert z-z^\prime \vert}$. Here, $k$ denotes the effective longitudinal wave vector of the nanoguide mode. Following this approach, the rate of spontaneous emission into the mode of a nanoguide reads
\begin{equation}
\Gamma_\mathrm{ng}= \frac{3c}{4\pi \epsilon^{3/2} v_g } \frac{\lambda^2}{A_\mathrm{eff}}\Gamma_0\,,
\end{equation}
where $v_g $ is the group velocity of light in the nanoguide and $A_\mathrm{eff}$ denotes the effective mode area of the nanoguide. We note that this treatment implicitly includes emission into the two opposite nanoguide directions.

If the radius of curvature of the ring is significantly larger than the emission wavelength, diffraction can be neglected and the resonator be locally treated as a linear nanoguide \cite{Hümmer2013}. Hence, the transverse part of the Green's tensor for a ring resonator remains the same as that of a linear nanoguide while the longitudinal component is modified according to $G_\mathrm{1D,cav}=\sum_{n=0}^{\infty} (e^{ikL})^n\frac{i}{2k} \Big( e^{ik\vert z-z^\prime \vert}+e^{ik(L-\vert z-z^\prime \vert)}\Big)$, taking into account all clockwise (CW) and counter-clockwise (CCW) partial waves for a ring of circumference $L$ \cite{Hümmer2013}. On resonance and for sufficiently large finesse ($\mathcal{F}$), the rate of emission for a dipole becomes 
\begin{equation}
\Gamma_\mathrm{cav}= 2\,\frac{\mathcal{F}}{\pi}\,\Gamma_\mathrm{ng}\,.
\label{finesse}
\end{equation}
Here, we use the conventional definition $\mathcal{F}=\frac{\mathrm{FSR}}{\mathrm{\Delta\nu}}$, where FSR denotes the free spectral range of the resonator, and $\Delta\nu$ stands for the full width at half-maximum (FWHM) of its resonance spectrum. Therefore, the enhancement of emission into the ring resonator is summarized by the quantity $\frac{\mathcal{F}}{\pi}$ since all other geometrical and material factors are included in $\Gamma_\mathrm{ng}$. It can be readily shown that this formulation is equivalent to the familiar $Q/V$ factor in Eq.\,(\ref{Purcell}) since the finesse, representing the number of times that a photon circulates in the resonator, incorporates the circumference of the resonating structure and the transverse area is accounted for in $\Gamma_\mathrm{ng}$. 

The factor of 2 in Eq.\,(\ref{finesse}) signifies the contributions of the two degenerate modes CW and CCW of a ring resonator. It is well known that these two modes can be coupled by a nanoscopic scatterer\,\cite{Mazzei2007}, whereby a line splitting occurs if the rate of scattering into the resonator mode becomes comparable with its loss rate. In our work, the rate $g$ of the coupling between a molecule and the resonator mode amounts to a few hundred MHz, which is considerably lower than the resonator linewidth of several tens of GHz. Hence, the resonator spectrum is not split. Instead, we expect individual molecules to cause narrow dips on the broad resonator line.

The power of a light field traversing a nanoguide changes from $P_{\rm 0}$ to $P \, =\,P_{\rm 0}(1-\beta_{\rm ng})^2\,$ after encountering a single emitter \cite{Faez2014}, where the coupling of the latter to the nanoguide is characterized by $\beta_{\rm ng} = \Gamma_{\rm ng} / (\Gamma_{\rm ng}+\Gamma_{\rm out})$, with $\Gamma_{\rm out}$ signifying the rate of emission into free space. When the emitter is coupled to a ring resonator, the coupling parameter becomes 
\begin{equation}
\beta_{\rm cav}=\frac{\Gamma_{\rm cav}}{\Gamma_{\rm cav}+\Gamma_{\rm out}}=\frac{\frac{2\,\mathcal{F}}{\pi}\,\beta_{\rm ng}}{\frac{2\,\mathcal{F}}{\pi}\,\beta_{\rm ng}+1}\,,
\label{beta}
\end{equation}
whereby both CW and CCW modes are accounted for\,\cite{Aoki2006,Srinivasan2007,Junge2013}. It, thus, follows that Purcell enhancement (see Eq.\,(\ref{finesse})) improves $\beta$, resulting in a stronger extinction effect by a single molecule. An important consequence of this enhancement is that it can compensate for intrinsically low coupling efficiencies caused by the molecular branching ratio or imperfect position and alignment of the emitter dipole moment. However, one should also bear in mind that because light scattered in the counter-propagating mode cannot interfere with the mode of the incident field in the drop or transmission ports, the dip and peak associated with the molecule are reduced at these ports. The new expression reads $P\,= \,P_{\rm 0}(1-\beta_{\rm cav}/2)^2$, yielding $P/P_{\rm 0}=25\%$ for the ideal case of $\beta_{\rm cav}=1$. We note that because the molecular ZPL has a linear polarization, the presence of a longitudinal electric field\,\cite{Junge2013} in our nanoguides does not affect our analysis.

\section{Experimental methods}
\begin{figure}[h]
	\center
	\includegraphics[width=0.8\linewidth]{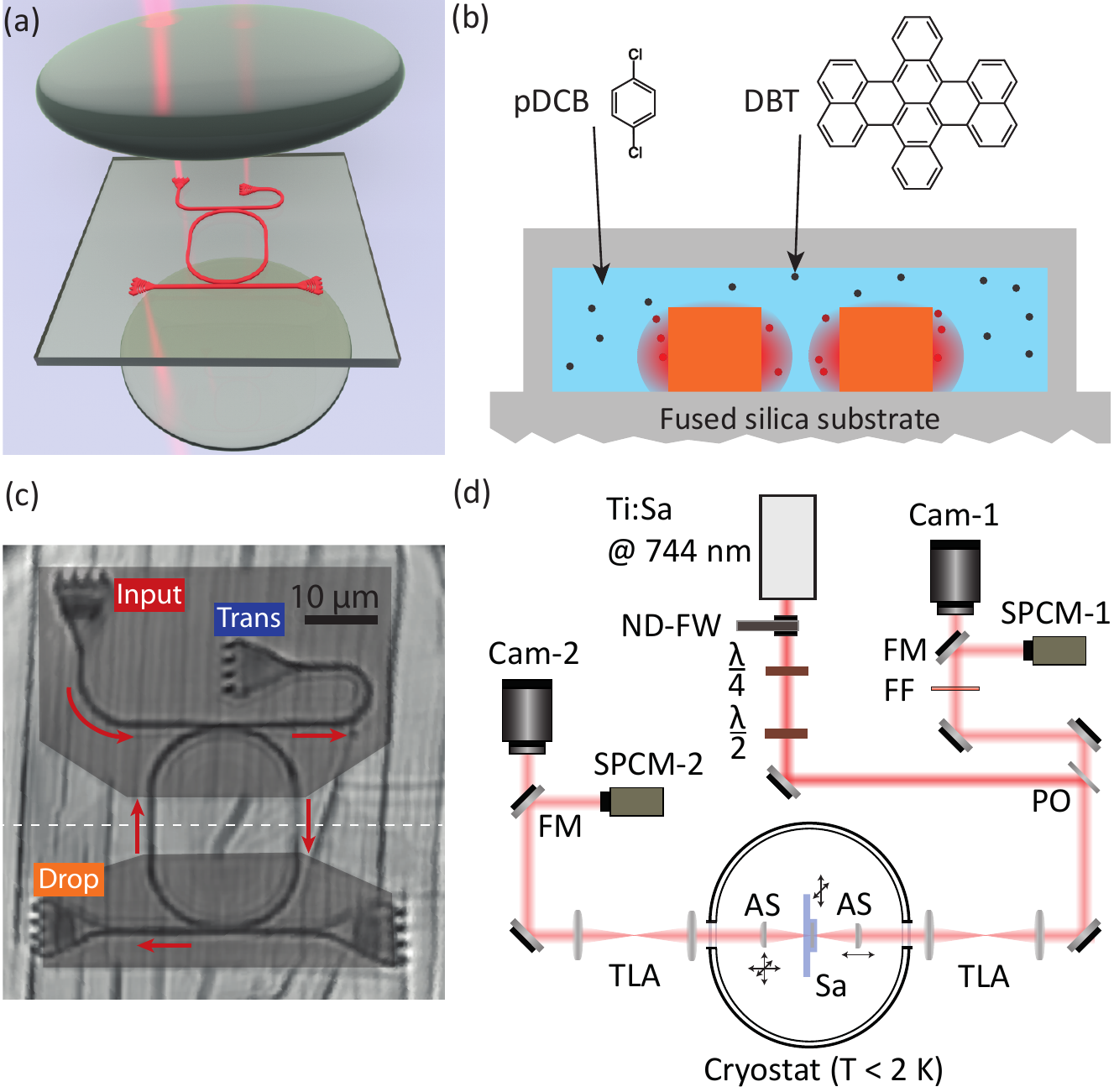}
	\caption{(a) Schematics of the microresonator arrangement with two aspheric lenses. (b) Schematic cross section of our nanoguide-resonator design along the dashed line in (b). The $\mathrm{TiO}_2$ nanoguide (refractive index $\mathrm{n}=2.4$) is surrounded by the doped pDCB matrix ($\mathrm{n}=1.5$), which fills the microchannels of fused silica ($\mathrm{n}=1.46$). The relative dimensions are not to scale. The structural formulas of pDCB and DBT are shown above the cross section. (c) Wide-field image of a ring resonator that is partially covered by resist pads (dark grey), recorded at 2 K. The red arrows show the direction of propagation of the in-coupled light. Dark vertical lines are caused by crystal defects, which appear during cool down. (d) Schematics of the optical setup. Abbreviations are: Ti:Sa: Continuous-wave titanium sapphire laser; ND-FW: filter wheel with neutral density filters; $\frac{\lambda}{2}$: half-wave plate;  $\frac{\lambda}{4}$: quarter-wave plate; PO: beam pick-off; TLA: telecentric lens assembly; FM: flip mirror; SPCM: single-photon counting module; Cam: camera; FF: fluorescence filter; AS: aspheric lens; Sa: sample. }
	\label{schematics}
\end{figure}

We use microresonators made of ring-shaped nanoguides coupled to straight nanoguides on a glass chip, which serve as input and output ports, see figure \ref{schematics}(a,c). The nanoguides are fabricated from a homogeneous $\mathrm{TiO}_2$ layer using electron beam lithography and reactive ion etching \cite{Türschmann2017}. A second substrate with a pre-structured indented channel of microscopic depth (1.4 $\mu \mathrm{m}$) is bonded from the top. This channel is filled with a molten DBT-doped pDCB via side openings (see figure \ref{schematics}(b) for a cross section of the structure). Upon a controlled decrease of temperature the liquid solidifies and forms a crystal around the nanoguides \cite{Türschmann2017,Gmeiner2016}. The subwavelength dimensions of the nanoguides lead to strong evanescent tails, thus allowing molecules in the vicinity to interact with their modes. 

The finesse of a ring resonator is determined by its round trip losses, which can stem from several mechanisms. First, one has to consider the leakage due to evanescent coupling with the straight nanoguides. Second, scattering caused by surface roughness of the microresonator leads to propagation losses. Third, the crystalline structure of the molecular host matrix can take on nanoscopic and microscopic irregularities upon formation and in the cooling process, also leading to scattering losses. Furthermore, bending losses might become significant for small resonator radii of curvature. Absorption in $\mathrm{TiO}_2$, however, is negligible for our wavelength and range of finesse. 

\begin{figure}[h]
	\center
	\includegraphics[width=0.8\linewidth]{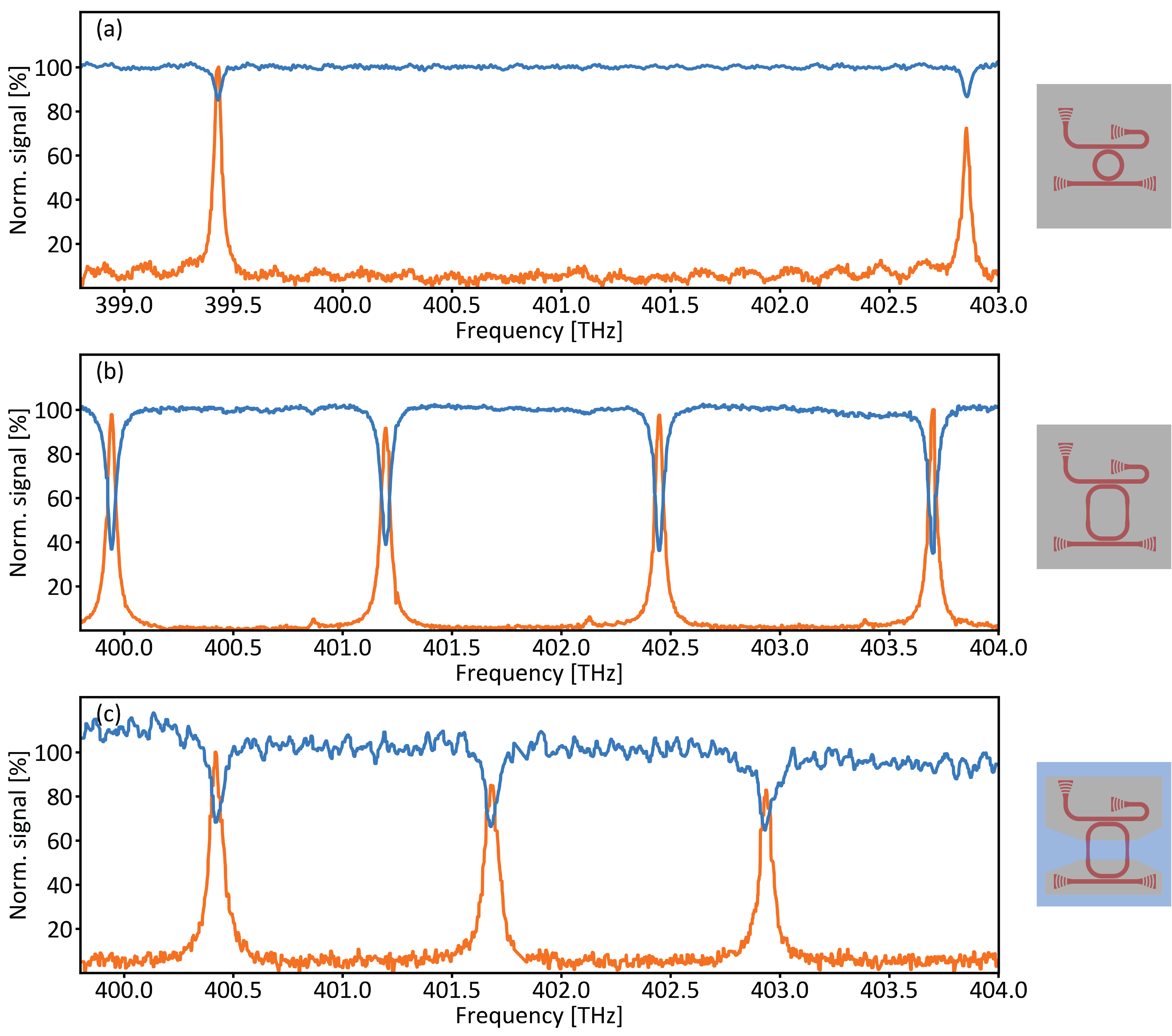}
	\caption{(a) Spectra recorded at room temperature from a circular resonator with radius of five micrometers covered with PMMA. Blue (orange) lines show the transmission (drop) port signals. The resonance shows $\Delta\nu=40$\,GHz and $\rm FSR=3.5$\,THz. (b) Spectra of a microresonator with identical design as in figure\,\ref{schematics}(c) but completely covered by PMMA and recorded at room temperature. The resonance shows $\Delta\nu=50$\,GHz and $\rm FSR=1.25$\,THz. (c) Spectra of the partially protected microresonator shown in figure\,\ref{schematics}(c). The spectrum was measured at 2 K and at the same excitation power as the molecule measurements in this work. The resonance shows $\Delta\nu=70$\,GHz and $\rm FSR=1.25$\,THz. The schematics on the right illustrate the resonator used in each case. Grey denotes PMMA or polymer resist. Blue represents pDCB.}
	\label{res-spectrum} 
\end{figure}

Figure\,\ref{res-spectrum}(a) displays an example of spectra recorded at room temperature from a circular microresonator with radius of 5\,$\mu$m covered with poly methyl methacrylate (PMMA). Here, light from a tunable continuous-wave laser was coupled through the upper left grating (see figure\,\ref{schematics}(a,c)) and read either through the transmission port (upper right corner), or the drop port (lower left corner). The signal at the transmission (drop) port shows dips (peaks) signifying cavity resonances with $\Delta\nu=40$\,GHz and $\rm FSR=3.5$\,THz, corresponding to $\mathcal{F} \approx 90$. 

As we shall see shortly, the finesse drops when the microresonator is covered with pDCB. In order to minimize the effect of crystal dislocations and cracks on the finesse, we employ a racetrack design and cover the curved parts by a polymer (mr-DWL micro resist technology GmbH, $\mathrm{n} \approx 1.5$) so that only the straight sections are exposed to pDCB, see figure \ref{schematics}(c). This minimizes the geometric mismatch of the crystalline axes with the resonator surface. In addition, the width of the nanoguide is widened from $300 \; \mathrm{nm}$ to  $400 \; \mathrm{nm}$ in the protected area in order to reduce the evanescent field and, thereby, scattering losses. Figure \ref{res-spectrum}(b) presents an example of room-temperature spectra recorded from a racetrack resonator fully covered with PMMA, displaying $\mathcal{F}=25$ with $\Delta\nu=50$\,GHz and $\rm FSR=1.25$\,THz. We attribute the lower finesse mainly to more scattering losses over the larger circumference of 82\,$\mu$m as compared to 31\,$\mu$m in figure\,\ref{res-spectrum}(a).

Figure \ref{res-spectrum}(c) shows the spectra from a racetrack resonator with identical design as that measured in figure\,\ref{res-spectrum}(b), but now partially-protected with resist, covered with pDCB and measured at 2\,K. We find that the resonator finesse is only slightly reduced to $\mathcal{F}=18$, which we attribute to residual small crystal defects. Unless stated otherwise, all low-temperature resonator data presented in this work were measured from this resonator. We note in passing that each fabricated chip contains a large number of resonator structures. Initial inspection of these allows one to select those with the best performance with regard to local crystal quality and molecule density.

The sample was optically accessed from both sides through two aspheric lenses with numerical aperture of 0.77, see figure \ref{schematics}(a,d). Light from a tunable Ti:Sapphire laser could be launched to and retrieved from these nanoguides via grating couplers. The outcoupled light was analyzed either by single-photon counting modules (SPCM-1/2, Excellitas Technologies SPCM-AQRM-14) or with cameras (Cam-1, Andor iXon 897; Cam-2, Hamamatsu ORCA-Flash 4.0). For fluorescence measurements bandpass filters (FF, Semrock ET785/60) were used. 

\section{Coherent coupling of molecules to a microresonator}
\begin{figure}[h]
	\center
	\includegraphics[width=0.8\linewidth]{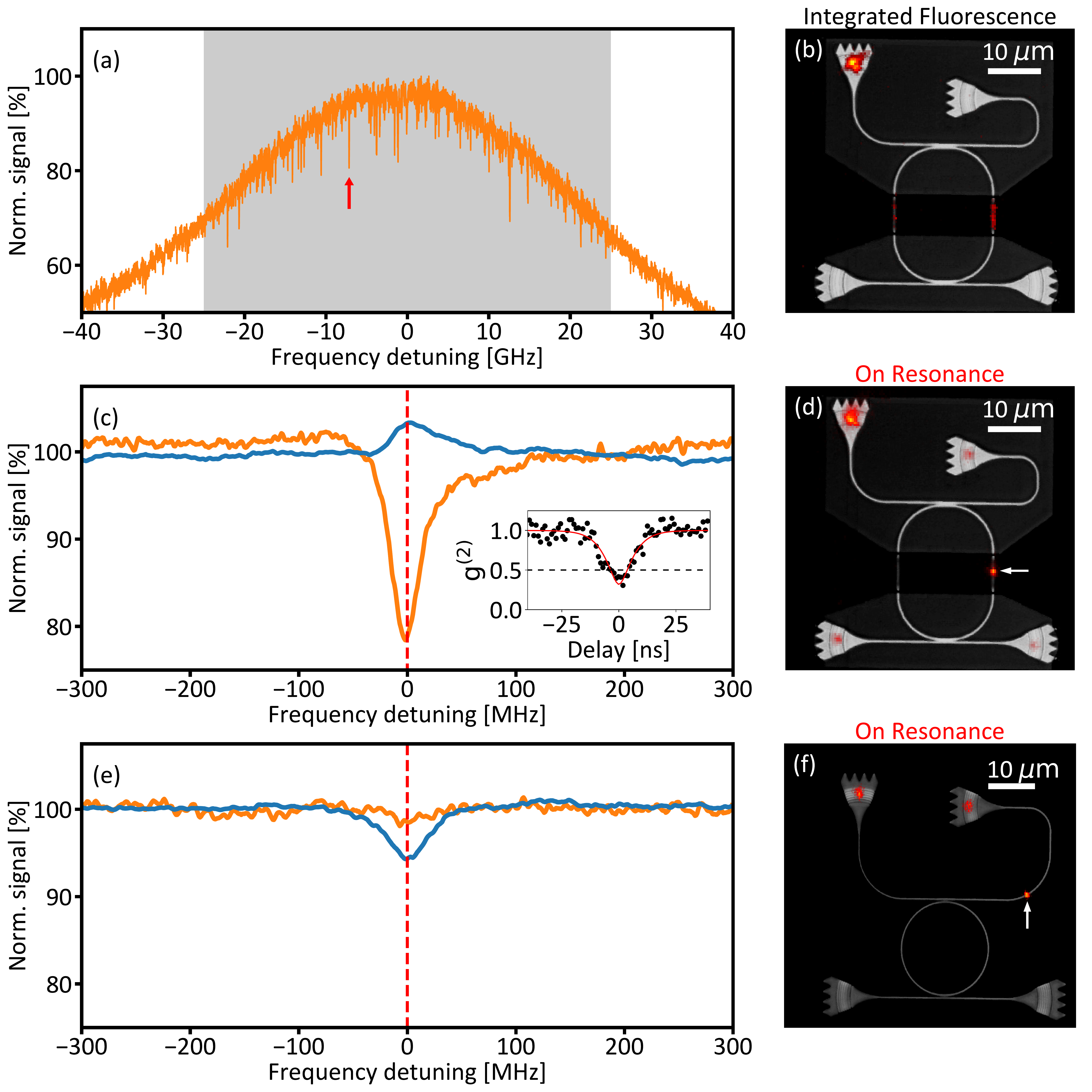}
	\caption{(a) Zoom into the resonance recorded at the drop port at 402.94 THz in figure\,\ref{res-spectrum}(c). A large number of narrow dips in the spectrum show coupling of molecules to the resonator. (b) Overlay of a bright-field image of the structure with an accumulated red-shifted fluorescence image of the resonator for the excitation frequency range highlighted in grey in (a). The two fluorescence strips along the exposed linear parts show that molecules are only excited there. (c) Averaged transmission (blue) and drop (orange) port signals of a single resonator-coupled molecule detuned by $-7 \; \mathrm{GHz}$ from the center of the resonance. This resonance is marked with a red arrow in (a). Inset shows the second-order autocorrelation of the molecular fluorescence. (d) Same as (b) but recorded at the ZPL of a single molecule indicated by the white arrow. (e),(f) same as (c),(d) but for a molecule (marked by the white arrow) which is coupled to the upper nanoguide of a circular resonator. Since this molecule is not coupled to the resonator, its signal is only visible in the transmission port. }
	\label{mol-extinction}
\end{figure}

Having identified the resonances of our microresonator, we now take a closer look at the resonance at 402.94\,THz of figure\,\ref{res-spectrum}(c). As shown in figure\,\ref{mol-extinction}(a), a frequency zoom into this resonance reveals narrow extinction dips caused by a large number of individual molecules coupled to the microresonator. The transitions of these chemically identical molecules undergo an inhomogeneous broadening caused by slight differences in their local environment. To ensure that each resonance can be attributed to a single molecule, in figure\,\ref{mol-extinction}(c) we zoom further by about one hundred fold onto one of the resonances marked by an arrow. Here, we performed repeated frequency scans and aligned the resonances in individual scans before averaging to account for slight spectral diffusion in the order of one linewidth. We find that the coherent interaction of the resonator mode with a single molecule leads to an extinction dip of 22\% in the drop port and a peak of 5\% in the transmission port. The peak appears weaker because the transmission through the undercoupled resonator alone amounts to a 25\% dip. Thus, the total effect of the molecule on the transmission channel follows $22\% \times 25\% \approx 5\%$. We note that the line shapes of the dip and the peak are not symmetric but resemble a Fano profile because the molecule was red-detuned from the cavity resonance by approximately $7\,\mathrm{GHz}$ ($10\%$ of the FWHM). Using the relation $P/\,P_{\rm 0}\,=1-0.22=0.78= (1-\beta_{\rm cav}/2)^2$ established earlier, we arrive at $\beta_{\rm cav}=23\%$.

The FWHM of about 34\,MHz for the resonance depicted in figure\,\ref{mol-extinction}(c) corresponds to the natural linewidth of DBT in pDCB and provides a strong evidence that the narrow resonances in figure\,\ref{mol-extinction}(a) stem from single molecules. To verify this further, we performed Hanbury Brown and Twiss measurements on the red-shifted fluorescence from the emitter. The observed antibunching with $g^{(2)}(0)=0.3$ displayed in the inset of figure\,\ref{mol-extinction}(c) clearly testifies to the single-molecule origin of the signal (the deviation from zero is consistent with the measurement of the background).

The ability of our setup to image the interaction region on the chip (see cameras Cam-1,2 in figure\,\ref{schematics}(d)) allows us to locate individual molecules with respect to the resonator and nanoguide architecture. Here, we again exploit the red-shifted fluorescence emitted out of the chip plane. Figure\,\ref{mol-extinction}(d) displays this for the incident light resonant with the ZPL of the single molecule studied in figure\,\ref{mol-extinction}(c), confirming that the data originate from a well-defined location next to the microresonator. In figure\,\ref{mol-extinction}(b), we show the cumulative fluorescence images recorded as the laser frequency was scanned over 50\,GHz accounting for 82 molecular resonances. Again, the image nicely shows that the molecules causing the extinction dips are all in the close vicinity of the exposed part of the microresonator. To contrast these results, in figure\,\ref{mol-extinction}(e,f) we also present an example of a molecule located along a linear nanoguide, i.e. not coupled to the resonator. As shown by the spectra in figure\,\ref{mol-extinction}(e), a weak extinction dip of about 5\% on the transmission port and no signal in the drop port report on the degree of coupling to a nanoguide without the cavity effect. 

\begin{figure}[h]
\center
	\includegraphics[width=0.8\linewidth]{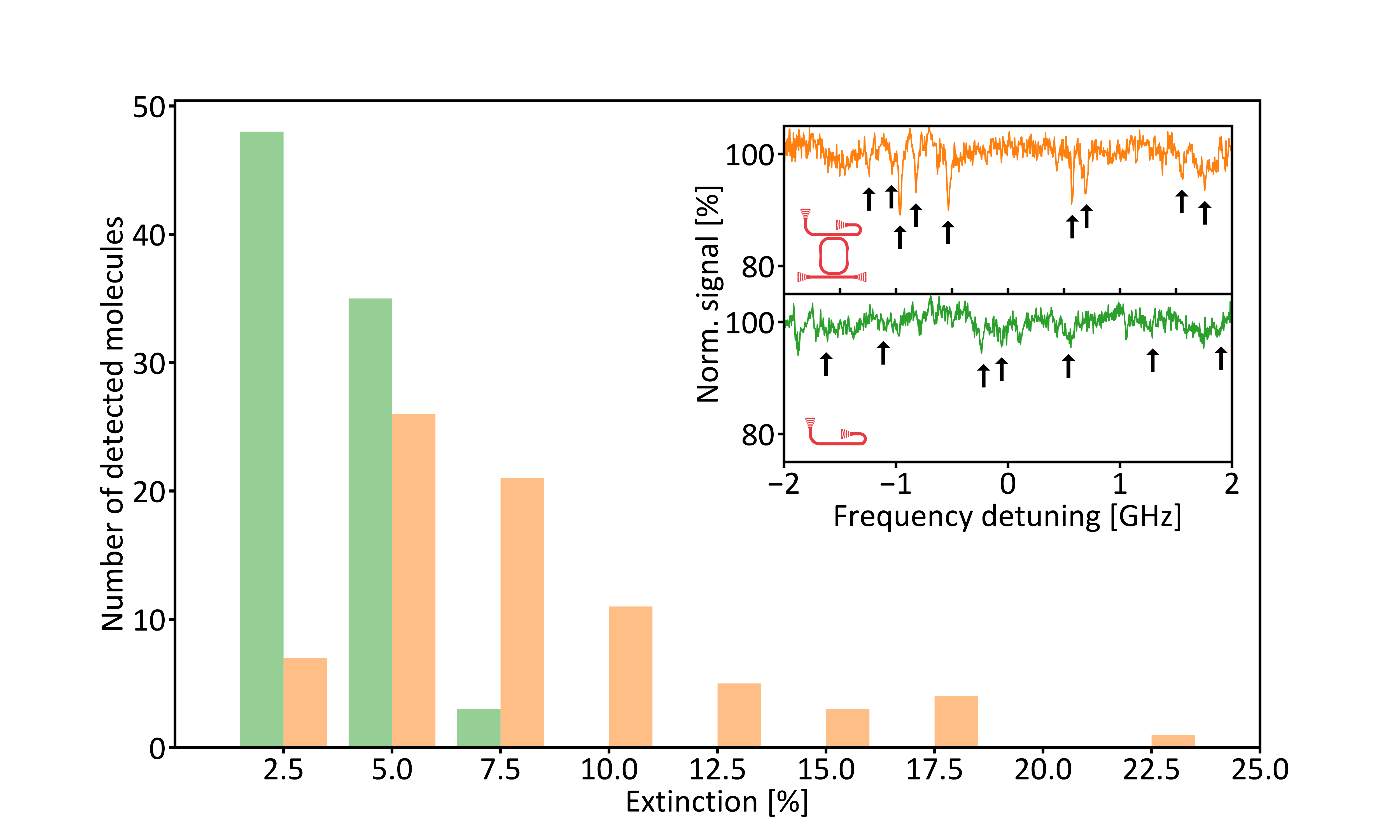}
	\caption{Histogram of extinction values recorded at the drop port of a racetrack resonator (orange) and from a linear nanoguide (green). The inset presents a section of the raw signals as well as sketches of the respective structures. The arrows in the inset point to examples of dips that were significant enough to be included in the analysis. Suitable molecules were detected via their red-shifted and out-scattered radiation using a peak finder algorithm. For the resonator data, molecules within the grey region of figure\,\ref{mol-extinction}(a) were evaluated. The histogram clearly shows larger extinction dips recorded from molecules coupled to the resonator. } 
	\label{histograms}
\end{figure}

The degree of coupling for individual molecules varies due to the spread in their distances and orientations with respect to the microresonator. Therefore, we studied the statistical distribution of the size of extinction dips from 82 molecules within the grey band shown in figure\,\ref{mol-extinction}(a). These data are presented by the orange histogram in figure\,\ref{histograms}, revealing extinction values reaching up to 22\%. To asses the effect of the microresonator enhancement, we compared our findings with the outcome of measurements on 87 molecules along a linear nanoguide with an identical cross-section but without partial polymer coating. Since the microresonator was detuned from the inhomogeneous band of DBT in this case, the presence of the microresonator could be ignored. The green histogram in figure\,\ref{histograms} summarizes the outcome of the measured extinction values for emitter-nanoguide coupling, reaching only about 7\%. Again, if we use the relation $P\, =\,P_{\rm 0}(1-\beta_{\rm ng})^2\,$ we can extract $\beta_{\rm ng}=3.5\%$ for the coupling to a simple nanoguide, which is much smaller than the maximum measured value of $\beta_{\rm cav}=23\%$ mentioned above. Thus, we arrive at about seven-fold enhancement, which is in good agreement with the eight-fold theoretically predicted enhancement for $\beta$ (see Eq.\,(\ref{beta})) if we consider $\mathcal{F}=18$ corresponding to a Purcell factor of 11. A clear separation of the two histograms in figure\,\ref{histograms} also provides a further strong evidence for the enhancement of the coherent coupling by our on-chip microresonators. 

\section{Outlook}

We have reported a seven-fold increase in the coupling efficiency of single organic molecules to a chip-based microresonator as compared to a linear nanoguide. Higher coupling efficiencies of $\beta_{\rm cav}\approx 90\%$ can be achieved if the current value of finesse were to be increased to $\mathcal{F}\approx500$ while $\beta_{\rm cav}\approx 95\%$ would require $\mathcal{F}=1000$. This will demand improvements in the microresonator quality as well as its coupling to molecular samples. As presented in figure\,\ref{res-spectrum}(a), our current fabrication process can deliver finesse in the order of 100, which we believe is limited by the etching quality of $\rm TiO_2$. In addition to improvements on this front, the quality of the molecular sample can be ameliorated by exploring different organic matrices and fabrication techniques, e.g. by using polymeric media \cite{Donley2000,Walser2009} or by placing individual nanocrystals\,\cite{Toninelli2018} next to the microresonator. 

We have demonstrated that a large number of molecules can couple to the same microcavity resonance. By integrating micro- and nano-electrodes on the chip architecture\,\cite{arXiv:1702.05923v1}, we plan to tune the resonance frequencies of the individual molecules at will, thus, setting the ground for the generation of polaritonic light-matter states and many-body quantum optical interactions. As compared to other material systems (atoms, quantum dots, color centers), the large density of molecules provides a unique platform for reaching these goals. 

We acknowledge the financial support by the Max Planck Society, an Alexander von Humboldt professorship and from the German Federal Ministry of Education and Research, co-funded by
the European Commission (project RouTe), project number 13N14839 within the research program "Photonik Forschung Deutschland". D.R. and A.S. contributed equally to this work. We thank Pierre T\"urschmann and Nir Rotenberg for their contributions to the early stages of this work. We are also grateful to the MPL micro- and nanofabrication team (TDSU-1) for continuous support.

\section*{References}
\bibliographystyle{iopart-num_mod}
\bibliography{bibliography}
\end{document}